\documentstyle[11pt,epsf]{article}    
\begin{document}   
  
\newcommand{\bra}{\langle}      
\newcommand{\ket}{\rangle}        
\def\tr{{\rm tr}\,}  
\def\href#1#2{#2}

\begin{titlepage}  
  
\begin{center}  
\hfill UOSTP-02101
\\  
\hfill hep-th/0204033 
\vspace{2cm}  
  
{\Large\bf  Supersymmetric Branes in the Matrix Model of
PP Wave Background}

\vspace{1.0cm}  
{\large  
Dongsu Bak   
}  
  
\vspace{0.6cm}  
  
{\it  
Physics Department, University of Seoul, Seoul 130-743, Korea  
\\ [.4cm]}  
({\tt dsbak@mach.uos.ac.kr 
})  
  
\end{center}  
\vspace{1.5cm}  
  
  
We consider the matrix model associated with pp-wave background and 
construct  supersymmetric  branes.  
In addition to the spherical membrane preserving 16  
supersymmetries, one may construct   rotating elliptic membranes   
preserving 8 supersymmetries.  
The other branch describes  rotating 1/8 BPS hyperbolic  
branes in general. When the angular momentum vanishes, 
the  hyperbolic brane becomes 1/4 BPS preserving 8 real  
supersymmetries.  
It may have the shape of hyperboloid of one or two sheets 
embedded in the flat
 three space.  
We study the spectrum of the worldvolume fields on the hyperbolic branes 
and  show that there are no  
massless degrees. We also compute the spectrum of the   
0-2 strings.   
 
\vspace{3.5cm}  
\begin{center}  
\today  
\end{center}  
\end{titlepage}  
  
\section{Introduction}

Recently there has been much attention to 
the string theories in the pp wave background, which may be  
obtained from the Penrose limit of $AdS_p\times S^q$  
geometries\cite{blau,
metsaev,maldacena,metsaev1,blau2,itzhaki,
gomis,russo,zayas,kim,cvetic,alishahiha,takayanagi,gursoy,
michelson,das,chu,dabholkar,berenstein,plee,
lu}. Remarkably the string spectrum in this background arises
from a certain subsector of $N=4$ super Yang-Mills 
theories\cite{maldacena} extending the usual AdS/CFT 
dualities.
 
The matrix model of the pp wave background is also  
constructed in Ref. \cite{maldacena}. The related  
pp wave metric may be obtained from the Penrose  
limit of $AdS_7\times S^4$,  
which arises as a near horizon limit of M5 branes. 
Compared to the usual matrix model\cite{BFSS}, 
the matrix model in the pp wave 
background  
includes mass terms and  couplings  to the  
background four form  
field strength.  
All the degrees become massive and the flat 
directions of the usual matrix model completely  
disappear, which makes the theory more accessible than the  
usual matrix model. Despite the presence of the mass term and couplings, 
 the model in the pp wave background still possesses real 32 
supersymmetries.  
 
Due to the Myers effect\cite{myers} 
induced by the background four form  
field strength, the matrix theory allows 1/2 BPS  
fuzzy sphere solutions\cite{maldacena}. The fuzzy membrane 
corresponds to the giant graviton wrapping two sphere of $S^4$ 
expanded by the effect of the four form flux\cite{toumbas}. 
Other simple time  
dependent 1/4 BPS configurations are constructed there, which may be 
interpreted as a collection of D0 branes rotating around a  
fixed axis with an angular frequency corresponding to the 
natural frequency provided by the harmonic potentials of  
mass term.  These are nothing to do with the expanded brane-like 
configurations.  
 
In this note, we  study more general BPS brane  
solutions in the matrix model. For the related discussions on the
D branes in pp wave background, see 
Refs. \cite{blau2,cvetic,chu,dabholkar,plee,lu}. 
The usual flat D2 branes cannot be  
supported in this model 
because all the coordinates acquire mass terms. 
Instead we find two branches of   
membrane solutions. The first describes 
rotating ellipsoidal branes preserving 
8 real supersymmetries. By the effect of the rotation, 
the fuzzy sphere gets deformed 
to the ellipsoidal shape. Contrary to  our expectation, 
 the lengths of three axes of the ellipsoid become all  
different and $SO(3)$ invariance is broken down to $Z_2$. 
Within this branch, if one increases the angular 
momentum further, the shape of brane becomes hyperbolic. 
The relevant BPS equations may be viewed as a deformation of
the BPS equations arising in the matrix model description of the 
supersymmetric tubes\cite{mateos,klee,swkim}.

The other branch in general describes a rotating  
hyperbolic membrane preserving 1/8 supersymmetries. 
The configuration is embedded in the three space formed by the two 
 coordinates  from $AdS_7$ and the remaining one  
along $S^4$ direction. When the angular momentum vanishes in  
this branch, the configuration becomes 1/4 BPS. The brane takes  
the shape of hyperboloid. It could either take 
the shape of upper/lower part of  two-sheet hyperboloid or 
 one-sheet hyperboloid ($dS_2$)  
embedded in the flat three space. 
We study the detailed solutions of the static  hyperbolic  
branes. The relevant solutions are provided by the unitary  
representation of the $SO(2,1)$ algebra. There are five class 
of unitary representation of $SO(2,1)$ and one may find 
the corresponding brane interpretation for 
 all of the classes. 
 
We further study the spectrum of the transverse scalar fields 
on the hyperbolic  branes. The supersymmetric 
 noncommutative solitons
on the branes describe D0 branes located at the origin. The
translational moduli  are completely lifted by the mass terms.
Using these solutions, we compute the spectrum of 0-2  
strings connecting D0 branes to the hyperbolic branes. 
 
In Section 2, we introduce the matrix model in 
the pp wave background and 
review the fuzzy membrane solutions and the  collectively 
rotating D0 brane solutions.
In Sections 3 and 4, 
we study the supersymmetry conditions introducing
relevant projection operators. In particular, we study the two classes of 
BPS states. We present the corresponding solutions including
rotating ellipsoidal membrane, the static hyperbolic branes and
the rotating hyperbolic branes.
In Section 5, we discuss the detailed solutions 
of  static hyperbolic branes using the unitary representation 
of SO(2,1) algebra and give the interpretation in terms of brane 
configurations.
We study the spectrum of the scalar field on the worldvolume of 
hyperbolic brane. Using the noncommutative solitons 
 on this brane, we investigate the spectrum of 0-2
strings.

\section{Matrix theory and Spherical membrane}                                                          
The matrix model in the pp-wave is constructed in  
Ref. \cite{maldacena}. The related geometry is $AdS_7 \times S^4$    
and, in the  
Penrose limit, the metric and the four form fields become 
\begin{eqnarray}    
&&ds^2= -4 d x^- dx^+ - 
 \left({\mu\over 6}\right)^2 \left( 4(x_1^2+x_2^2 
+x_3^2) +(x_4^2+\cdots + 
x_9^2)\right)  
(dx^+)^2 + dx^i dx^i\nonumber\\ 
&& F_{+ 123}= \mu 
\end{eqnarray}    
where $1,2,3$ directions were coordinates of $S^4$ and the remaining six  
directions are related to the spatial directions of the $AdS_7$. 
One of the angular directions in $S^4$ is used 
in the light cone coordinate. 
The matrix     
model  Lagrangian\cite{maldacena}  in this background becomes 
\begin{equation}   
L= L_0 + L_\mu       
\end{equation}    
with 
\begin{eqnarray}    
&&L_0={1\over 2 R}\tr \left(  \sum_i (D_0 X_i)^2     
+{R^2}
\sum_{i<j} [X_i,X_j]^2    
\right)  +\tr \left(\psi^{T} D0 \psi + i \sum_i 
R \psi^{T} \gamma_i [\psi, X_i] \right)\nonumber\\ 
&& L_\mu= 
-{1\over 2 R} \left({\mu\over 6}\right)^2\tr \left( 4\sum^3_{i=1}X_i^2  
+\sum^9_{i=4} X_i^2\right) 
-{\mu\over 4}\tr \psi^{T} \gamma_{123} \psi  
-{\mu\over 3}i\sum^3_{i,j,k=1}\!\!\!\!\tr X_i X_j X_k  
\epsilon_{ijk}\,,     
\label{lag}     
\end{eqnarray}    
where
 we set the eleven dimensional    
Planck length $l_{11}=1$.  $R=g_s \sqrt{\alpha'}$ is the     
radius of the direction $2x^-$ and  
$l_{11}$ is related to the string scale by  
$ l_{11}=( 2\pi \alpha' R)^{1\over 3}$. 
The 16 dimensional Majonara spinors are used 
for the fermionic part 
and the gamma matrices are taken to be real.  
The scale $R$ (together with  $l_{11}$)   
will be  omitted below by setting them unity.    
The Lagrangian $L_0$ is the same as the usual matrix model  
in \cite{BFSS}. $L_\mu$ includes mass terms and the coupling  
to the four form field background.   
 
The system possesses 32 real supersymmetries in total. The fields  
transform  
under the supersymmetry as
\begin{eqnarray}    
&&\delta X = 2 \psi^T \gamma _i \epsilon(t),\ \ \  
\delta A_0= 2 \psi^T \epsilon(t)\nonumber\\ 
&& \delta \psi ={1\over R}\left( D_0 X_i \gamma_i + 
{\mu\over 3}  
\sum^3_{i=1}X_i\gamma_i \gamma_{123}  
-{\mu\over 6}\sum^9_{i=4} X_i \gamma_i \gamma_{123} 
+{iR\over 2} [X_i,X_j] \gamma_{ij} 
\right)\epsilon(t)    
\end{eqnarray}    
with 
\begin{equation}     
\epsilon(t)= e^{-{\mu\over 12}\gamma_{123}t}\epsilon_0\,.  
\end{equation}  
For the remaining 16 supersymmetries, only fermions transform as  
$\delta \psi = 
e^{{\mu\over 4}\gamma_{123}t}\tilde{\epsilon}_0$ while all the bosonic  
coordinates do not change. Like the usual matrix model, this  
part of the supersymmetries are realized nonlinearly due to the  
presence of N D0 particles. 
 
Because of the mass terms, the flat directions of the usual matrix 
model completely disappear. Further the coupling term to the four  
form field strength  may induce the dielectric effect\cite{myers}.  
 
In this note we shall focus on the nonperturbative states in the matrix 
model of the  
pp wave background. As found in Ref. \cite{maldacena}, there are  
solutions preserving 1/2 supersymmetries. We briefly review  
this fuzzy sphere configuration for the later comparison.  
From the supersymmetric  
transformation of the fermionic coordinate, the BPS equation  
becomes 
\begin{equation}     
[X_i, X_j]= i {\mu\over 3} \epsilon_{ijk} X_k    
\end{equation}    
for $i,j,k=1,2,3$, $D_0 X_i=0$ for all $i$ and  
$X_4=\cdots= X_9=0$. The solutions  are given by fuzzy sphere; 
writing $X_i = {\mu\over 3} L_i$, one sees that 
\begin{equation}     
[L_i, L_j]= i \epsilon_{ijk} L_k    
\end{equation}    
and solutions are given by a unitary representation of $SU(2)$. 
For the $n=2j+1$ dimensional representation with Casimir $j(j+1)$, 
the fuzzy sphere is described by  
\begin{equation}     
X_1^2+X_2^2 + X_3^2=   \left( {\mu\over 3}\right)^2 j(j+1) 
=\left( {\mu\over 6}\right)^2 (n^2-1)\,. 
\end{equation}     
The translations 
do not give new solutions and the related moduli  are lifted  
completely.  Since $1,2,3$ coordinates are related to $S^4$, 
it is clear that above configuration wraps a sphere in $S^4$ 
and may be identified a giant graviton in the pp-wave  
background\cite{toumbas}. 
 
In Ref. \cite{maldacena},  solutions preserving 1/4 of  
supersymmetries were found. An example is 
\begin{eqnarray}     
(X_4+ i X_5) (t)= e^{\pm i {\mu\over 6}t} (X_4+ i X_5) (0)\,,\ \ \ \   
   [X_4,X_5]=0\,. 
\label{bpst}     
\end{eqnarray}       
In the basis diagonalizing $X_4$ and $X_5$ simultaneously,  
D0 branes taking definite time dependent  positions rotate 
around the origin at the same angular frequency as their 
natural frequency $\mu/6$ of the harmonic potential.  
Hence we see that the motion  
of D0 branes are the usual particle motion in the  
harmonic potential, which is nothing to do with the formation of 
 expanded higher dimensional  
branes.   
 
Of course, similar solutions may be obtained by replacing 
(4,5) by any other pairs. In addition, there is a similar 1/4 
BPS solution by taking pair of indices from 1,2,3 with time 
dependence $e^{\pm i{\mu\over 3}t}$. Again the angular frequency 
and the natural frequency agree to each other. 
 
\section{Rotating Ellipsoidal Branes} 
 
In this section we shall investigate more closely 
the case where only the first three components of $X_i$ 
are turned on. The resulting configuration in general preserves 
1/4 supersymmetries restoring 1/2 supersymmetry 
in the static limit. 
As we will see, the configuration  corresponds to a 
deformation of the spherical  
membrane to a time dependent ellipsoidal one. There is at  
least one parameter family of solutions 
describing the deformation. If this parameter 
approaches a certain critical value, the ellipsoid collapses  
in two directions. Beyond the critical value, the membrane surface 
takes a shape of hyperboloid. 
 
From the combination of $\gamma_i\,\, (i=1,2,3)$, the only real  
projection operator  is of the type 
\begin{eqnarray}  
P_\pm^1={1\pm \hat{n}_i \gamma_i\over 2}\,, 
\end{eqnarray}  
where $\hat{n}_i$ is a real vector of unit size. Without loss of  
generality, we shall choose  $\hat{n}_i$ to the $3$ 
direction using the global $SO(3)$ symmetry.  
 
 The supersymmetry condition, 
 $\delta \psi=0$, then leads to BPS equations, 
\begin{eqnarray}  
&& i[X_1, X_2]  + {\mu\over 3} X_3=0 \,, \ \  
D_0 X_3 =0\nonumber\\ 
&& i[X_1, X_3](\pm)  -   
{\mu\over 3} X_2 (\pm) + D_0 X_1=0 \nonumber\\ 
&&  i[X_2, X_3](\pm)   +{\mu\over 3} X_1 (\pm)+ D_0 X_2=0\,,  
\label{bpse} 
\end{eqnarray}  
by setting the coefficient of $P^1_\pm \epsilon_0$. 
For the remaining supersymmetries, we shall satisfy the equations 
for just one choice of the sign.  
In addition, there is the Gauss law 
constraint, 
\begin{eqnarray}  
[X_1, D_0 X_1]+ [X_2, D_0 X_2]=0\,. 
\end{eqnarray}  
For definiteness, we shall choose $+$ sign projection  
corresponding to the remaining  
supersymmetries satisfying  
$\epsilon_0= - \gamma_3 \epsilon_0$. Working in a gauge  
$A_0= X_3$, the last two equations reduce to 
\begin{eqnarray}  
\dot{X}_1 +i\dot{X}_2 = -{\mu\over 3}i (X_1+i X_2) \,, 
\end{eqnarray}  
whose solution reads 
\begin{eqnarray}  
X_1+i X_2 = e^{-{\mu\over 3}it} (X_0+i Y_0) \, 
\label{time} 
\end{eqnarray}  
with constant Hermitian matrices $X_0$ and $Y_0$.  
The second equation in (\ref{bpse}) implies that $X_3\equiv Z$ is  
constant in time. 
Using this solution, 
the full set of BPS equations are reduced to 
\begin{eqnarray}  
[X_0, [X_0, Z]]+ [Y_0, [Y_0, Z]] -2\left({\mu\over 3}\right)^2  
Z=0\,,\ \ \  
[X_0, Y_0]  = i{\mu\over 3} Z \,. 
\label{bpsre} 
\end{eqnarray}  
These   generalize  the BPS equations associated with  
the supersymmetric tubular branes by a mass  
parameter\cite{klee,swkim}. 
The general solutions of these coupled equations are not known. 
 
When $Z=0$, the above equations may be trivially satisfied and  
the solutions in (\ref{time}) with commuting $X_0$ and $Y_0$ describe 
the rotating  1/4 BPS D0 branes discussed in the last  
section. 
 
For more general solutions, we try the following ansatz, 
\begin{eqnarray}  
[Z, X_0]  = i{\mu\over 3}(1+a)Y_0 \,,\ \ \  
[Y_0, Z]  = i{\mu\over 3}(1-b)X_0\,.  
\label{ansatz} 
\end{eqnarray}  
The first equation in (\ref{bpsre}) then implies that $a=b$. 
The momentum  becomes 
\begin{eqnarray}  
D_0 X _1 +iD_0 X_2 = -i{\mu a\over 3} e^{-{\mu\over 3}it} 
(X_0-i Y_0) \,, 
\end{eqnarray}

When $|a| \le 1$, the solution may be presented as 
\begin{eqnarray}  
 X_0={\mu\over 3}\sqrt{1+a} L_1 \,,\ \ \  
Y_0={\mu\over 3}\sqrt{1-a}L_2\,,\ \ \ 
Z= {\mu\over 3} \sqrt{1-a^2}L_3 
\label{sole} 
\end{eqnarray}  
with the generators of the SO(3) algebra  
$[L_i,L_j]=i\epsilon_{ijk} L_k$. 
The shape is determined by the Casimir of the SO(3) representation 
and given by 
\begin{equation}     
{X_0^2\over 1+a}+{Y_0^2\over 1-a} + {X_3^2\over 1-a^2}  =    
\left( {\mu\over 3}\right)^2 j(j+1)\,. 
\label{ellipse}     
\end{equation} 
 
When $a=0$, the solution becomes spherical and $D_0 X_1=D_0 X_2=0$. 
Though we are using here a different gauge, this corresponds to 
 the 1/2 BPS 
spherical membrane described in the previous section. 
Otherwise the solutions describe rotating ellipsoidal branes, which is 
1/4 BPS. 
The total energy for this configuration and the angular momentum 
are evaluated as 
\begin{eqnarray}  
&& H= \left({\mu a \over 3}\right)^2 \tr (X_0^2 +Y_0^2) 
={ a^2\over 6} \left({\mu  \over 3}\right)^4 n (n^2-1)\,,\nonumber\\ 
&& J_{12}= \tr (X_1 D_0 X_2 -X_2 D_0 X_1)= - 
{ a^2\over 6} \left({\mu  \over 3}\right)^3 n (n^2-1) 
\end{eqnarray}  
for the $n$ dimensional representation of the SO(3) algebra. 
As we see from (\ref{ellipse}), the rotation in the 12 plane makes 
the shape of the brane elliptical.  However the deformation
is more than expected. The lengths of three axes become
all different and $SO(3)$ R-symmetry of the model is 
broken down to at most discrete subgroups.

For $|a| > 1$, the rotating membrane takes the shape of  
hyperboloid. Specifically for $a>1$,   
 the solution may be presented as 
\begin{eqnarray}  
 X_0={\mu\over 3}\sqrt{1+a} K_1 \,,\ \ \  
Y_0={\mu\over 3}\sqrt{a-1}K_2\,,\ \ \ 
Z= {\mu\over 3} \sqrt{a^2-1}K_3 
\end{eqnarray}  
with the generators of the SO(2,1) algebra  
$[K_1,K_2]=i K_3$, $[K_3,K_1]= i K_2$ and  
$[K_2,K_3]=- i K_1$.  The Casimir operator in this case 
is $K=K_2^2+K^2_3-K_1^2$ and the shape is described by 
\begin{equation}     
{X_0^2\over 1+a}-{Y_0^2\over a-1} -{X_3^2\over a^2-1} =-  
\left( {\mu\over 3}\right)^2 K\,. 
\end{equation}  
 
There is a  nonvanishing  angular momentum  in the 
$12$ plane. 
The detailed representation of the SO(2,1) algebra  
will be presented  
in the next section where we discuss the  static 
hyperboloid.

\section{Rotating 1/8 BPS Hyperbolic Branes} 
To look for the other BPS configurations, let us turn on  
one component out of 1,2,3 and two components out of the remaining 
6 directions. Specifically we turn on $X_3$, $ X_4$ and $X_5$. 
The unbroken supersymmetry condition, 
 $\delta \psi=0$, may be written as 
\begin{eqnarray}     
&&\ \  M_+\left( D_0 X_4 \gamma_4 + D_0 X_5 \gamma_5 - 
{\mu\over 6}(X_4 \gamma_4+X_5 \gamma_5) \gamma_{123}  
+i[X_4,X_3] \gamma_{43}+ 
i[X_5,X_3] \gamma_{53} 
\right)\epsilon_0  \nonumber\\  
&&+ M_-\left( D_0 X_3 \gamma_3 +  
{\mu\over 3}X_3 \gamma_3 \gamma_{123}  
+i [X_4,X_5] \gamma_{45} 
\right)\epsilon_0 =0\,, 
\end{eqnarray}     
where $M_\pm\equiv e^{\pm{\mu\over 12}\gamma_{123}t}$. 
We now introduce real projection operators  
\begin{eqnarray}  
P_\pm^1={1\pm \gamma_3\over 2}\,,\ \ \ 
P_\pm^5={1\pm \gamma_{12345}\over 2}  
\end{eqnarray}  
and it is straightforward to see that  the above equation reduces 
to 
\begin{eqnarray}  
&& i[X_4, X_5](+) (\pm)  =   {\mu\over 3} X_3 (\pm)(+)\,, \ \  
D_0 X_3 =0\nonumber\\ 
&& i[X_4, X_3](\pm) (+)  =    
{\mu\over 6} X_5 (+)(\pm)- D_0 X_4 \nonumber\\ 
&&  i[X_5, X_3](\pm) (+)  = -{\mu\over 6} X_4 (+)(\pm)- D_0 X_5  
\label{bpsh} 
\end{eqnarray}  
by setting coefficients of $P^1_\pm P^5_\pm\epsilon_0$ to zero. 
Here the signatures in the first parenthesis of each term 
represent that of the projection operator $P^1_\pm$ 
whereas the second signatures for $P_\pm^5$. $(+)$ represents that 
the projection operators are not involved. 
The configuration should satisfy the Gauss law, 
\begin{eqnarray}  
[X_4, D_0 X_4]+ [X_5, D_0 X_5]=0\,. 
\end{eqnarray}  
 
For the remaining  
supersymmetries, the above equations should be satisfied at least 
one set of the choice of the signatures.  Let us choose the  
$P^1_+ P^5_- $ case. The other choices can be treated similarly. 
When  $Z=0$, $[X_4, X_5]=0$ follows from the first equation of  
(\ref{bpsh}).  
The remaining equation reduce to the 1/4 BPS 
equation (\ref{bpst}) discussed in the last section with the  
remaining supersymmetries   given by         
$P_\pm^5 \epsilon_0$. 
 
Non rotating  solutions 
are given only if $D_0 X_4= D_0 X_5=0$. The corresponding BPS  
equations become  
\begin{eqnarray}  
 [X_4, X_5] = +i{\mu\over 3} X_3 \,,\ \  
 [X_5, X_3] = -i{\mu\over 6} X_4 \,, \ \  
 [X_3, X_4] = -i{\mu\over 6} X_5 \,.  
\label{bpshh} 
\end{eqnarray}  
The configuration is 1/4 BPS and the remaining supersymmetries are 
given by the projection  
$(P^1_+ P^5_- + P^1_- P^5_+ )\epsilon_0$. This projection is equivalent to 
a single projection operator $P^4_-$ defined by
\begin{eqnarray}  
P^4_\pm \equiv {1\pm \gamma_{1245}\over 2}\,. 
\end{eqnarray}  
One could use  
$P^1_+ P^5_+ + P^1_- P^5_- =P^4_+$ by which again we get SO(2,1) 
algebra corresponding to parity operation of (\ref{bpshh}). 
Later we shall analyze details of this static case.

For a more general case,  again we choose a gauge $A_0=X_3$ and  
 the projection $P^1_+ P^5_-$. The last two BPS equations  
then become 
\begin{eqnarray}  
\dot{X}_4 +i\dot{X}_5 = {\mu\over 6}i (X_4+i X_5) \,, 
\end{eqnarray}  
whose solution reads 
\begin{eqnarray}  
X_4+i X_5 = e^{{\mu\over 6}it} (X_0+i Y_0) \, 
\label{timeh} 
\end{eqnarray}  
with constant Hermitian matrices $X_0$ and $Y_0$.  
The second equation in (\ref{bpse}) imply that $X_3\equiv Z$ is  
constant in time. 
Using this solution, 
the BPS equations are reduced to 
\begin{eqnarray}  
[X_0, [X_0, Z]]+ [Y_0, [Y_0, Z]] +\left({\mu\over 3}\right)^2  
Z=0\,,\ \ \  
[X_0, Y_0]  = i{\mu\over 3} Z \,. 
\label{bpsreh} 
\end{eqnarray}  
The essential difference from (\ref{bpsre}) is the signature 
in front of the mass square term. These equations in general  
describe 
1/8 BPS configurations preserving 4 real supersymmetries. 
Again we do not know how to solve the equation in general. 
Taking the  ansatz, 
\begin{eqnarray}  
[Z, X_0]  = -i{\mu\over 6}(1+a)Y_0 \,,\ \ \  
[Y_0, Z]  = -i{\mu\over 6}(1-b)X_0\,,  
\label{ansatzh} 
\end{eqnarray}  
one finds $a=b$ from  (\ref{bpsreh}). 
The momentum in 45 directions then becomes 
\begin{eqnarray}  
D_0 X _4 +iD_0 X_5 = i{\mu a\over 6} 
(X_0-i Y_0)e^{{\mu\over 6}it}  \,. 
\end{eqnarray}  
The $a=0$ case corresponds to the 1/4 BPS configuration with 
vanishing  momentum. For  general $a$, the configuration 
takes a shape 
\begin{equation}     
{X_0^2\over 2(1+a)}+{Y_0^2\over 2(1-a)}  - {X_3^2\over 1-a^2}  =    
\left( {\mu\over 6}\right)^2 {\rm Casimir}\,, 
\label{shape}     
\end{equation} 
rotating in 45 plane. 
 
The total energy  and the angular momentum 
for this 1/8 BPS configuration 
are evaluated as 
\begin{eqnarray}  
&& H 
=\left({ \mu\over 6}\right)^2 \tr \left( (a^2-a+1) X_0^2+ (a^2+a+1)Y_0^2 
+4 Z^2\right) 
\,,\nonumber\\ 
&& J_{45}= \tr (X_4 D_0 X_5 -X_5 D_0 X_4)=  
{ a\mu \over 6} \tr ( X_0^2- Y_0^2)\,. 
\end{eqnarray}  
As $|a|$ grows, the angular momentum in general increases and the shape  
changes accordingly. But in this 1/8 BPS branch, the instant shape 
of the brane configuration takes always the form of hyperboloid as we  
seen in (\ref{shape}).

\section{Static Hyperbolic Branes and Their Fluctuation Spectrum} 
 
In the last section, we have obtained a static 1/4 BPS 
configurations.  
With definitions 
\begin{eqnarray}  
 X_4={\mu\over 3\sqrt{2}} K_1 \,,\ \ \  
X_5={\mu\over 3\sqrt{2}}K_2\,,\ \ \ 
Z= {\mu\over 6} K_3\,, 
\end{eqnarray}  
the   BPS equations in (\ref{bpshh}) implies  
that $K_i$  satisfies the SO(2,1) algebra. Namely, 
$[K_1,K_2]=i K_3$, $[K_3,K_1]= -i K_2$ and  
$[K_2,K_3]=-i K_1$, which we write  
as $[K_i,K_j]=i f^k_{ij} K_k$ using a SO(2,1)  
structure constant $f_{ij}^k$.  
The Casimir operator in this case 
is $K=K_1^2+K^2_2-K_3^2$ and the shape of the configuration  
 is described by 
\begin{equation}     
{1\over 2}({X_4^2}+{X_5^2}) - {X_3^2} =  
\left( {\mu\over 6}\right)^2 K\,. 
\label{hyperboloid}     
\end{equation}  
 
The BPS branes are classified by all possible unitary 
representations of $SO(2,1)$ algebra, which have been already 
worked out in literatures\cite{dixon,ooguri,polychronakos}. 
The basic idea is to introduce the step operators 
by 
\begin{equation}     
K_\pm = K_1 \mp i K_2\,, 
\end{equation}  
which satisfy the commutation relations 
\begin{equation}     
[K_3,K_\pm]=\pm K_\pm\,,\ \ \ [K_-, K_+]=2 K_3\,. 
\end{equation} 
Let us work in a basis  diagonalizing $K$ and $K_3$, 
the eigenvalues of which are denoted respectively by 
$k$ and $m$.  
Using $K_-= K_+^\dagger$ in the unitary representation,  
the relation 
\begin{equation}     
K_\mp K_\pm= K + K_3(K_3\pm 1)\,. 
\end{equation}   
gives a requirement 
\begin{equation}     
k + m(m\pm 1)\ge 0\,, 
\end{equation} 
for all $m$. Then there are basically the following  
five class of unitary  
representations parametrized by $j$ with $k=-j(j-1)$.

(1) Discrete representations ${\cal D}_j^+$ 
 
The representation  is realized in the Hilbert space 
\begin{equation}     
{\cal D}_j^+=\{|jm\ket; m=j,j+1,j+2,j+3 \cdots\}\,. 
\end{equation}   
where $|jj\ket$ is annihilated by $K_-$. 
$j$ is real and positive for 
the unitary representation. 
 
(2) Discrete representations ${\cal D}_j^-$ 
 
The Hilbert space is 
\begin{equation}     
{\cal D}_j^-=\{|jm\ket; m=-j,-j-1,-j-2, \cdots\}\,. 
\end{equation}   
where $|j\,-\!j\ket$ is annihilated by $K_+$.  
$j$ is real and positive for 
the unitary representation. 
 
(3) Continuous representations ${\cal C}_j^\alpha$ 

The Hilbert space is 
\begin{equation}     
{\cal C}_j^\alpha=\{|\alpha;jm\ket; m=\alpha,\alpha\pm 1, 
\alpha\pm 2, \cdots\}\,, 
\end{equation}   
and $0\le \alpha  < 1$ without loss of generality. 
The representation is unitary if  
$j={1/2}+ is$ with real $s$. 
 
(4) Complementary continuous  
representations ${\cal E}_j^\alpha$ 
 
The Hilbert space is 
\begin{equation}     
{\cal E}_j^\alpha=\{|\alpha;jm\ket; m=\alpha,\alpha\pm 1, 
\alpha\pm 2, \cdots\}\,, 
\end{equation}   
and $0\le \alpha  < 1$ without loss of generality. 
For $ 1/2 < j  < 1 $ with $j-1/2 < |\alpha -1/2|$, 
the representation becomes unitary.  
  
(5) Identity representation 
 
This is trivial representation with $j=m=0$.

All of the nontrivial  unitary representations are solutions 
describing 1/4 BPS configurations.  
The case ${\cal D}_j^+$ describes a hyperbolic plane corresponding  
 to 
the upper plane of two-sheet hyperboloid as depicted in Fig. 1. 
There is a rotational symmetry in $45$ plane. 
In this case, the Casimir operator takes a value 
$k= 1/4 -(j-1/2)^2$ with $j > 0$. 
For $j\ge 1$, the upper part of plane defined by  
(\ref{hyperboloid}) precisely agrees to the one in Figure 1. 
When $0<j < 1$, the plane defined by (\ref{hyperboloid}) 
takes a shape of $dS_2$. Because the eigenvalue  of $X_3$ 
is bounded below in this case, we view that the brane 
still takes the shape of the upper plane of two-sheet hyperboloid 
in the  
figure even for $0<j \le 1$. Presumably the effect of  
noncommutativity on the brane induces this kind of  correction  
and makes the tip closed off. 
 
One may understand the configuration as follows. 
Without the pp wave background, there is a usual M2 brane  
in 45-plane that shares one direction 
with M5 branes.  
Now consider the situation where one holds the circular boundary of  
large M2 brane at fixed $X_3$ location. 
The mass term  in presence of the pp wave  
corresponds  harmonic confining potentials in the target space.  
Since the 
region at the origin has a lower potential, the M2 brane 
gets deformed toward the origin in order to take advantage  
of the lower potential. Of course, there is a cost of energy by 
the increase 
of area together with curving of the brane. 
These contributions are balanced in the above configurations.  
 
The other discrete representation ${\cal D}_j^-$ corresponds to the  
lower 
part of two-sheet hyperboloid. 
From ${\cal D}_j^+$, one may obtain ${\cal D}_j^-$ 
by the transformation $X_3\rightarrow -X_3$.

\begin{figure}[tb] 
\epsfxsize=2.7in 
\hspace{.8in}\epsffile{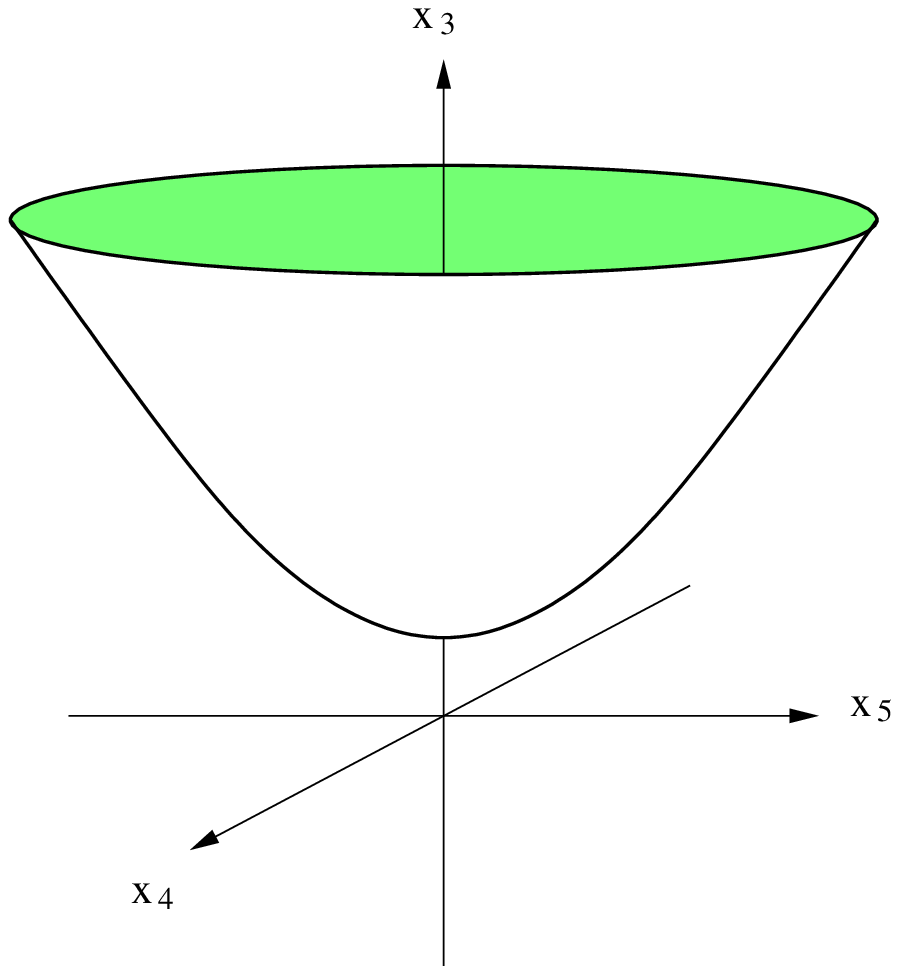} 
\vspace{.2in} 
\\ 
{\small Figure~1:~We illustrate here the upper plane of 
two-sheet hyperboloid.} 
\end{figure}

The continuous representation corresponds to a  
brane of the shape $dS_2$. Namely the upper  and  lower parts of 
two-sheet hyperboloid are connected by a tubular throat. The tube, 
in genral, 
does not close off along the throat. Also the D0 branes are located along  
the $X_3$ direction with equal spacing $\Delta= {\mu/6}$, which one can  
read off from the eigenvalues of  
$X_3$. For ${\cal C}^\alpha_j$, $k=1/4 + s^2$, which determines the 
classical shape.   
Especially, the fatness of the throat is determined by the Casimir.  
Unlike the case of the  
discrete representations, there is an extra parameter $\alpha$ for a  
given $k$.  
Since $D0$ branes in the $X_3$ directions are located with the equal  
spacing $\Delta$, the configuration has a periodicity $\Delta$ 
along $X_3$ directions about the locations of $D0's$.  
Hence the parameter $\mu\alpha/6$ $(0\le \alpha < 1)$ precisely  
describes the locations of a D0 within $0 \le x_3 < \Delta$.

For ${\cal E}^\alpha_j$, $k =1/4- (j-1/2)^2 \in (0,1/4)$ with $ 1/2 <j <1$ 
and the throat becomes narrower as $j$ approaches  
$1$. 
The effect of the noncommutativity  
along the throat becomes crucial in this case. 
In particular, the freedom of locationg $D0$ brane in the unit interval 
gets lifted  due to the too narrow  
throat. And the lattice translational freedom of $D0$ brane location 
breaks down partially. 
By this one may understand the ristriction of the locational  
parameter $\alpha$ given by $j-1/2 < |\alpha-1/2 |$. 
The detailed dynamical investigation of the $k$ dependence of 
the D0 brane location will be quite interesting.

Finally we like to discuss the spectrum of the fluctuation 
around the hyperbolic brane background. For simplicity,  
we like to take 
the case of the upper part of two-sheet hyperboloid 
of the representation  
${\cal D}_j^+$.  
Among the degrees, we like to concentrate on the transverse  
fluctuation along  $6,7,8,9$. Let us call them $\phi$. Then
the effective Lagrangian governing the quadratic fluctuation  
becomes 
\begin{equation}     
L_{\rm eff} = {1\over 2}\tr \left(\dot\phi^2 - 
\left({\mu\over 6}\right)^2 \phi^2 -\sum_{i=3,4,5}\phi [\bar{X}_i, 
[\bar{X}_i,\phi]] \right) \,, 
\label{eff}     
\end{equation}  
where $\bar{X}_i$ denotes the brane solution. Here  
we choose a  
gauge $A_0=0$.  In order to find the spectrum, one has to  
diagonalize the above effective action, for which we proceed as  
follows.  
The scalar field in general may be written in components as 
\begin{equation}     
\phi =\sum_{nm}\phi_{nm} |jn\ket\bra jm| \,. 
\end{equation}  
We now define   
\begin{equation}     
 \tilde{K}_i\phi \equiv -\phi K_i \,, 
\end{equation}  
which acts on $\bra jm|$. Then it is simple to show that  
they satisfy the SO(2,1) algebra, i.e. 
$[\tilde{K}_i, \tilde{K}_j]=i f^k_{ij}\tilde{K}_k$. 
Hence, 
\begin{equation}     
 [K_i, \phi]= (K_i+\tilde{K}_i)\phi= J_i \phi    
\end{equation}  
with  the total generator $J_i= K_i+\tilde{K}_i$.  
Also note that the bases $\bra jm|$ form the representation  
${\cal D}^-_j$ with respect to $\tilde{K}_i$. 
The problem is then reduced to the decomposition of the product  
representation of ${\cal D}^+_j\otimes {\cal D}^-_j$ into sum of  
irreducible representation of the $SO(2,1)$ algebra. 
They are given by\cite{polychronakos} 
\begin{equation}     
{\cal D}^+_j\otimes {\cal D}^-_j = \int^\infty_{s=0}  
{\cal C}^{\alpha=0}_{1/2+is} \,.
\end{equation} 
Hence in these bases denoted by $\phi_{(s,n)}$ with nonnegative  $s$ and  
integer $n$, one has 
\begin{equation}     
\sum_{i=3,4,5} [\bar{X}_i, 
[\bar{X}_i,\phi_{(s,n)}]]= \left({\mu\over 6}\right)^2  
\left({1\over 2}+2s^2  
+3 n^2\right)\phi_{(s,n)} 
\end{equation}  
By adding the contribution from the original mass term in (\ref{eff}) 
which is already diagonal in the basis $\phi_{(s,n)}$, 
the total mass squared  is 
\begin{equation}     
M^2_{s,n}=\left({\mu\over 6}\right)^2  
\left({3\over 2}+2s^2 + 3n^2\right)\,.   
\end{equation}  
Similarly one may compute the mass spectrum of the scalar in  
(1,2) directions and they are 
\begin{equation}     
\tilde{M}^2_{s,n}=\left({\mu\over 6}\right)^2  
\left({9\over 2}+2s^2 + 3n^2\right)\,.   
\end{equation}  
From the expressions, it is clear that there are no massless modes 
and there are no moduli related to the brane position in the 
 transverse space. 
 
Finally we like to mention the spectrum of 0-2 strings connecting 
the upper  brane of  two-sheet hyperboloid and D0 branes sitting at  
the origin. The D0 branes may be given as a localized  
noncommutative soliton solution\cite{bak,aganagic,harvey,park}  
from the view point  
of the worldvolume theory of  
the  brane. For simplicity, we consider the representation  
${\cal D}_j^+$. 
It is straightforward  to check the following 
solution satisfy the BPS algebra (\ref{bpshh}); 
\begin{eqnarray}  
 X_4={\mu\over 3\sqrt{2}} S K_1 S^\dagger \,,\ \ \  
X_5={\mu\over 3\sqrt{2}} S K_2 S^\dagger\,,\ \ \ 
Z= {\mu\over 6}S K_3 S^\dagger\,, 
\end{eqnarray} 
where we define a shift operator $S$ such that 
\begin{eqnarray}  
 S^\dagger S=I\,,\ \  S S^\dagger = I-P 
\end{eqnarray} 
with $l$ dimensional  projection operator $P$. 
The solution describes $l$ D0 branes sitting at  
the origin\cite{aganagic,park}. Again due to the mass terms in  
the model, the translational moduli of D0 branes completely  
disappear. In this BPS configuration, they are located at the  
origin, which is the minimum point of the  
potential.  More explicitly, for  
the projection operator $P=|jj\ket\bra jj|$, 
the corresponding shift operator may be constructed 
as 
\begin{eqnarray}  
 S= \sum^\infty_{n=0} |j,j\!+\!n\!+\!1\ket\bra j,j\!+\!n|\,. 
\end{eqnarray} 
In the background of this solution, one may compute the spectrum  
related to the 0-2 string. In particular  
$\phi$ in the transverse directions decomposed as  
$\phi= P\phi P  + P\phi \bar{P} + \bar{P} \phi P +  
\bar{P}\phi \bar{P}$ with $\bar{P}=I-P$. 
The quadratic part of each term contributes to the action  
independently. One may then explicitly check that $P\phi P$  
describes the fluctuation of D0's while $\bar{P}\phi \bar{P}$ 
part describes the fluctuation of the original  
hyperbolic brane. The remaining part is to do with the 0-2  
strings. Namely they describe the interaction between 
the D0 branes and the hyperbolic brane. In the transverse  
direction, the  spectrum may be evaluated  
straightforwardly. (See Refs. \cite{aganagic,park}  
for the detailed methods.) The resulting spectrum 
for $P\phi \bar{P}$ in the 6,7,8,9 directions 
reads 
\begin{eqnarray}  
 M_{na}^2= \left({\mu\over 6}\right)^2  
\left(1+ 2j(1-j) + 3(j+n)^2\right)\,, 
\end{eqnarray} 
where $a$ labels the a-th D particle and $n=0,1,2,\cdots$. 
For 1,2 direction, the mass spectrum is given by 
\begin{eqnarray}  
 \tilde{M}_{na}^2= \left({\mu\over 6}\right)^2  
\left(4+ 2j(1-j) + 3(j+n)^2\right)\,. 
\end{eqnarray} 
In these expressions, the constant parts , 
$1\times (\mu/6)^2$ or
$4\times (\mu/6)^2$, come from the original mass terms of the
Lagrangian (\ref{lag}). The remaining parts take the same form
and may be interpreted as the distance squared between
D0's at the origin and the n-th D0 brane on the hyperbolic brane.
More explicitly, the distance may be measured as
$d^2_n= \bra j\,j\!+\!n|(X_4^2+ X_5^2+ X_3^2)|j\,j\!+\!n\ket$ on the solution, 
which agrees with the 
remaining parts of the masses squared.

\section{Conclusion}  
 
In this note, we study the generic two brane solutions 
arising in the matrix model of the pp wave background. 
There are two branches we have identified. One is the 
1/4 BPS branch where the rotating branes take an ellipsoidal  
shape. The others include  1/4 BPS static hyperbolic branes
and rotating hyperbolic branes preserving
1/8 of the supersymmetries.
We focus on the static  branes of the hyperbolic shape and
find all possible solutions classified by the
unitary representation of the SO(2,1) algebra.
We have computed the spectrum of the transverse fluctuation
as well as 0-2 strings.

In the Matrix model of the pp wave background, 
it is clear that
 the static flat branes are not supported because of
mass terms. Also, the supersymmetric charges 
in this system 
 are time dependent and do not commute with the 
Hamiltonian, which makes it difficult to apply the 
conventional methods of finding BPS equations.
In this respect, we do not know the whole
 classifications of  possible BPS equations in 
the system. 
Also for the BPS equations  identified in this note, 
the general 
solutions are not known and  our solutions  are obtained
by taking specific 
ansatz. A more systematic understanding
of the BPS states in this massive matrix model 
is necessary.

The detailed dynamical understanding of the formation 
or deformation 
of the branes are still lacking.
For the rotating ellipsoidal shape of branes, the lengths of all 
the axes becomes different. 
If one increase the angular velocity,
the ellipsoidal brane opens up and becomes hyperbolic.
Why these are so dynamically is 
not clear to us. 
The further investigation  is
necessary. 

We have computed the spectrums of  the 0-2 string connecting D0 to 
the static 
hyperbolic brane. This  may be directly 
compared with those of string theory in the 
pp wave background.

\noindent{\large\bf Acknowledgment}   
We would like to thank Kimyeong Lee  for  
enlightening discussions and Nobuyoshi Ohta for comments..      
This work is supported in part by KOSEF 1998     
Interdisciplinary Research Grant 98-07-02-07-01-5.


\end{document}